# Phosphorus-Controlled Nanoepitaxy in the Asymmetric Growth of GaAs–InP Core–Shell Bent Nanowires

*Spencer McDermott, Trevor R. Smith, Ray R. LaPierre, and Ryan B. Lewis\**

Department of Engineering Physics, McMaster University, L8S 4L7 Hamilton, Canada

\*Email: rlewis@mcmaster.ca

**Abstract**

Breakthroughs extending nanostructure engineering beyond what is possible with current fabrication techniques will be crucial for enabling next-generation nanotechnologies. Nanoepitaxy of strain-engineered bent nanowire heterostructures presents a promising platform for realizing bottom-up and scalable fabrication of nanowire devices. The synthesis of these structures requires the selective asymmetric deposition of lattice-mismatched shells—a complex growth process which is not well understood. We present the nanoepitaxial growth of GaAs–InP core–shell bent nanowires and connecting nanowire pairs to form nano-arches. Compositional analysis of nanowire cross-sections reveals the critical role of adatom diffusion in the nanoepitaxial growth process, which leads to two distinct growth regimes: indium-diffusion limited growth and phosphorous-limited growth. The highly controllable phosphorous-limited growth mode is employed to synthesize connected nanowire pairs and quantify the role of flux shadowing on the shell growth process. These results provide important insight into three-dimensional nanoepitaxy and enable new possibilities for nanowire device fabrication.

The epitaxial growth of three-dimensional nano-heterostructures presents a vast design landscape to realize novel and creative nanostructures and devices with bottom-up and scalable fabrication. To harness these wide-ranging design opportunities, the complex three-dimensional (3D) deposition geometries and the role of adatom diffusion on faceted nanostructures present growth challenges that must be understood. Recently, spontaneous bending of free-standing nanowires with an asymmetric lattice-mismatched core–shell heterostructure have gathered interest, presenting novel strain and geometry engineering opportunities with applications in sensing and optoelectronics. These structures undergo spontaneous bending to relieve misfit strain between the core and asymmetric shell, which can be fabricated by directional deposition (selective flux exposure around the nanowire). Bent nanowires have been synthesized using molecular beam epitaxy (MBE),[1–5] metal−organic MBE,[6] and electron beam evaporation.[7–9] A variety of shell materials have been explored, including group III–V,[2–6,8,10–14] IV,[15] nitrides[16] and transition metal-based shells with III–V or IV cores.[1,7,9] Additionally, bent nanowires[17] and nanowires connected through bending[2,3] have been proposed as a scalable fabrication approach for ultra-sensitive sensors. InP-based nanowires have been used as transducers in field-effect transistors (FET)[18,19] and FET-based devices fabricated by bending nanowire pairs together has been proposed.[3]

For III-V nanowires, deliberate bending was shown to take place along the group-III flux direction, and the role of adatom diffusion has thus far been ignored.[2,3] In general, the distribution of the asymmetric shell determines the bending direction, and for positive lattice-mismatched shells, the nanowires bend away from where the shell forms.[2,3,12–14] However, recent reports by Al-Humaidi et al.[4,5] observed both bending along the V ($As_4$) flux direction and along the III (Ga) flux direction during the growth of the $In_xGa_{1-x}As$ shells on GaAs cores, although an explanation for this observation was not provided. Additionally, for the Bi surfactant-directed growth of InAs quantum dots on nanowire sidewalls, InAs growth was shown to occur on As-facing facets and not on In-facing facets.[2] These results highlight the need for a more detailed understanding of this nanoepitaxial growth process.

For GaAs MBE on planar and faceted GaAs surfaces, differences in the partial pressure of arsenic have been shown to alter the Ga incorporation diffusion length, driving selective and asymmetrical growth. For growth on faceted GaAs surfaces, the transfer of Ga adatoms (and growth) to facets receiving higher incident $As_4$ flux has been demonstrated.[20] The effect of arsenic partial pressure on Ga adatom incorporation diffusion length has been shown to be linear at lower and quadratic at higher arsenic pressures,[20,21] for both $As_4$ and $As_2$.[22] Similarly, InAs quantum dot growth on rippled GaAs surfaces are known to favor areas of the surface with locally higher incident $As_4$ flux.[23] The effect was observed at temperatures above 500 °C where In adatom diffusion was sufficient to enable selective growth.[23–26] However, the impact of adatom diffusion and incorporation for nanowire shell growth has not been explored.

In this work, we reveal the crucial role that adatom diffusion and deposition geometry play in the MBE growth of GaAs–InP core–shell bent nanowires and connected bent nanowire pairs. By varying the InP shell growth temperatures—and thus the In adatom migration length—the growth transitions from In-diffusion-controlled to phosphorous-flux-controlled, with the resulting InP shell geometry determined by the incident In and $P_2$ fluxes, respectively. Transmission electron microscopy (TEM) and energy dispersive x-ray spectroscopy (EDS) analysis elucidate the nanowire cross-sections and shell distributions, revealing the phosphorous-controlled growth regime as a stable and deterministic process for precise synthesis of bent nanowire structures. This growth regime is employed to synthesize bent nanowire pairs, which are of high interest for sensing applications. TEM and EDS analysis of these structures quantifies the impact of flux-shadowing and demonstrates that the connected nanowires are intimately fused together by the InP shell. These results will pave the way for the fabrication of bottom-up scalable nanosensors.

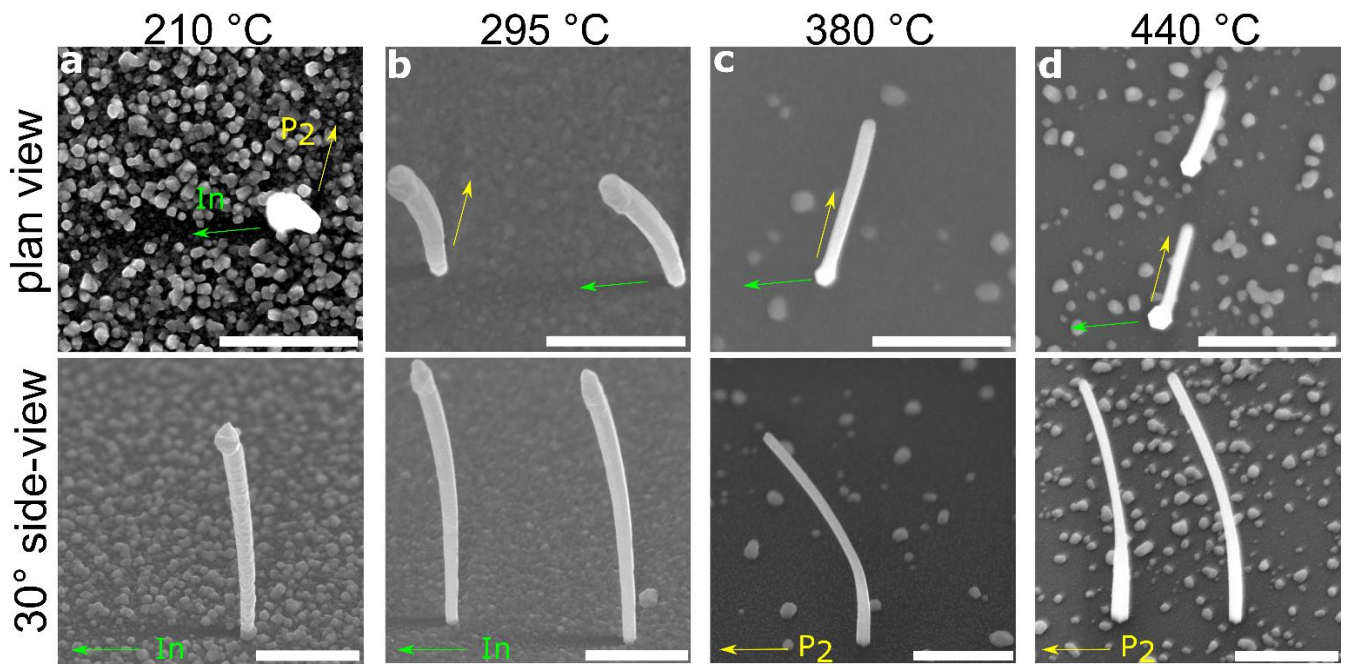

**Figure 1.** SEM images of GaAs-InP core–shell nanowires grown at InP shell growth temperature of 210 °C (a), 295 °C (b), 380 °C (c) and 440 °C (d). The top row shows plan view micrographs, indicating the bending direction with respect to the incident In and P$_2$ fluxes. The bottom row presents inclined side-view images aligned azimuthally perpendicular to either the incident In (a–b) or P$_2$ (c–d) flux. The In and P$_2$ source fluxes projected on the substrate are indicated by green and yellow arrows, respectively. Scale bars correspond to 1 µm.

Top-view and side-view SEM images of GaAs–InP core-shell nanowires grown with various InP shell growth temperatures are presented in Figure 1a–d. InP shell growth at the lowest temperature—210 °C—with a planar deposition of 40 nm (Figure 1a) exhibits little bending. We note that the nanowire sidewalls appear rough and there is substantial parasitic growth on the substrate at this growth temperature. Increasing the shell growth temperature to 295 °C at the same planar deposition (Figure 1b) results in smoother nanowire sidewall facets and more bending. The bending direction from the In flux is ~70° at the nanowire base, curling towards ~45° at the tip, suggesting that shell growth occurred predominantly on the facet with the highest overlapping In and P$_2$ fluxes. For higher growth temperatures of 380 °C and 440 °C with a planar deposition of 9 nm—Figures 1c and 1d, respectively—the nanowires are highly bent along the incident P$_2$ flux direction, suggesting that shell growth occurred predominantly on the P$_2$-facing facets. The most bending occurred at 380 °C with a 1.2 µm projected in-plane length. We note that Al-Humaidi et al. recently observed the bending direction of GaAs–In$_x$Ga$_{1-x}$As core–shell nanowires depended on the substrate.[4] The present findings suggest a difference in adatom diffusion along the nanowires as a possible explanation—from temperature or other factors.

Nanowires were characterized by cross-sectional TEM and EDS to further investigate the impact of substrate temperature on InP shell growth. Figure 2 presents high-angle annular dark field (HAADF) micrographs and EDS maps of nanowires grown at 210 °C with a planar deposition of 40 nm (a–c) and 380 °C with a planar deposition of 9 nm (d–f).

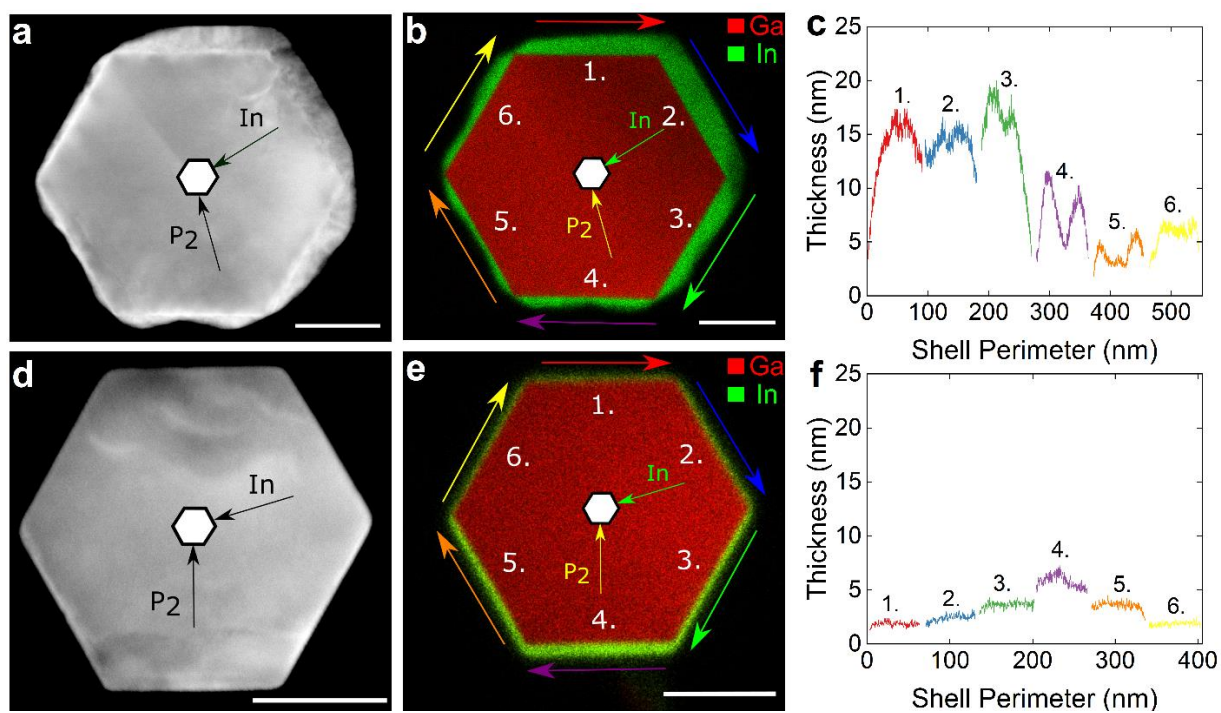

**Figure 2.** TEM investigation of microtome cross-sections for nanowires with shells grown at 210 °C (a–c) and 380 °C (d–f) presenting HAADF micrographs (a,d) and EDS maps (b,e). The EDS maps show that while shell growth occurred on all facets, it occurred predominantly on In-facing facets at 210 °C and $P_2$-facing facets at 380 °C. This is confirmed by plotting the shell thickness—extracted from the EDS maps—around the nanowire core, as illustrated in (c) and (f) for 210 °C and 380 °C, respectively. The incident flux directions are indicated in (a) and (d),

corresponding to In impingement on facet 2 and P$_2$ impingement on facet 4, respectively. The color and numbering of the EDS line scans corresponds to the labels on the EDS maps, indicating the path around the nanowire shell. Scale bars are 50 nm.

For shell growth at 210 °C, the HAADF image (Figure 2a) indicates irregular shell growth. The shell morphology and HAADF contrast is believed to be the result of plastic strain relaxation. We note that the formation of similar plastically-relaxed (In,Ga)As mounds was reported by Lewis et al.[27] The presence of plastic relaxation in the core–shell heterostructure explains why these nanowires do not exhibit significant bending, despite considerable asymmetry in the shell. The corresponding EDS map of In and Ga is shown in Figure 2b, demarcating the InP shell and GaAs core. The EDS map demonstrates that the facets with direct In impingement received the most InP deposition at 210 °C. EDS line scans around the shell are shown in Figure 2c, where the number and color corresponds to the labels in Figure 2b. We note the presence of contrast variations between shell facets in the EDS map and plotted thickness for growth at 210 °C (see Supporting Information). The three facets which received direct In impingement (line scans 1–3) all show more InP growth than the facets which did not receive direct In flux (line scans 4–6). The cross-sectional shell area was $1330\pm90$ nm$^2$ for facet 2, almost twice the $730 \pm 50$ nm$^2$ for facet 4. These results indicate that the diffusion of In around the nanowire was an important limiting factor in the shell formation.

In contrast to shell growth at 210 °C, the HAADF image of the nanowire grown at 380 °C (Figure 2d) exhibits a smooth hexagonal sidewall surface with a consistent contrast. In this case, the EDS map (Figure 2e) and thickness measurements (Figure 2f) show the three facets facing toward the P$_2$ flux all have thicker shells than those facing away. Specifically, the shell thicknesses on facets

3 and 5 are similar, despite facet 3 being exposed to the In beam and facet 5 facing away from the In flux—both facets received the same direct $P_2$ flux. The favoring of shell formation under the $P_2$ flux at 380 °C is similar to the selective growth of InAs QDs on wavy surfaces, where the QDs formed on areas with locally higher direct As impingement.[23] The results suggest InP asymmetric shell growth requires higher temperatures that have significant In adatom diffusion—unlike what has been shown for $In_xAl_{x-1}As$[2] and $In_xGa_{x-1}As$.[4]

If the adatom diffusion length is considerably larger than the nanowire circumference, we would expect the relative growth on each facet to be proportional to the In incorporation diffusion length. In planar GaAs growth studies, the Ga incorporation diffusion length was found to be linearly proportional to the impinging $As_2$ flux.[22] As the group-V surface diffusion length is negligible,[28] we expect the relative growth rate to be proportional to the incident $P_2$ flux on a facet in the high-In-diffusion limit. The sources of impinging phosphorus on the nanowire facets are direct impingement and scattering/reemission from the oxide mask. Assuming the scattered flux to be equivalent in all directions (equally scattered on all side facets) and assuming that the growth rate is directly proportional to the incident $P_2$ flux, the growth rate on a side facet is:

$$\frac{\partial g_f}{\partial t} = C\left[F_{P,direct}\langle \hat{b}_p \cdot \hat{n}\rangle + F_{P,scattered}\right]$$

where $F_{P,scattered}$ and $F_{P,direct}$ are the $P_2$ impingement from scattering and the direct beam, respectively. $\hat{b}_p$ is a vector pointing toward the $P_2$ source, $\hat{n}$ is the normal vector of the side facet and $C$ is a constant relating $P_2$ impingement to growth. From the average measured thickness of the side facets, we calculate growth rate contributions for the direct beam ($C\,F_{P,direct}\langle \hat{b}_p \cdot \hat{n}\rangle$) of

$0.16 \pm 0.03$ µm/h and scattering ($CF_{P,scattered}$) of $0.05 \pm 0.01$ µm/h (see Supporting Information). This corresponds to a $P_2$ scattering flux of $31\pm8\%$ of the total $P_2$ flux on the side facet 4. Given the flux orientation illustrated in Figure 2d, this corresponds approximately to a shell thickness ratio (and thus $P_2$ flux ratio) of 3:2:1 on facets 4:(3 and 5):(1,2 and 6) in Figure 2e,f. We note that Mohammed et al. [29] and Küpers et al. [30] reported a similar contribution from scattered As flux incident on isolated GaAs nanowires during MBE growth. The close agreement with the measured shell growth around the nanowire supports the assumption of an In diffusion length considerably larger than the nanowire cross-sectional dimension. However, we expect that as the group V flux increases, the diffusion length of In adatoms will decrease.[20–22] If the In diffusion length becomes comparable to or smaller than the nanowire circumference, the growth will begin to favor the In-facing facets, as is the case at 210 °C. We note that the projected flux angles will change throughout the growth as the nanowire bends [3]—which will have some effect on the incident fluxes.

Pair shadowing occurs when one nanowire blocks a unidirectional flux from reaching its neighbor. Recently, we reported group III shadowing effects in III-V nanowires.[3] In that work, we modeled shadowing for a perfectly aligned unidirectional beam. To explore the shadowing effect in the context of the $P_2$-mediated InP shell growth revealed above, growths were carried out with the nanowire pairs aligned along the $P_2$ beam. Shadowing also provides a means to vary the incident group V flux distribution around the nanowire cross-section and thus validate the above conclusions about the growth process. Figure 3a presents SEM images of nanowire pairs with varying spacing, grown with incident $P_2$ flux from the right, resulting in the partial shadowing of the direct $P_2$ flux for the left-hand nanowires.

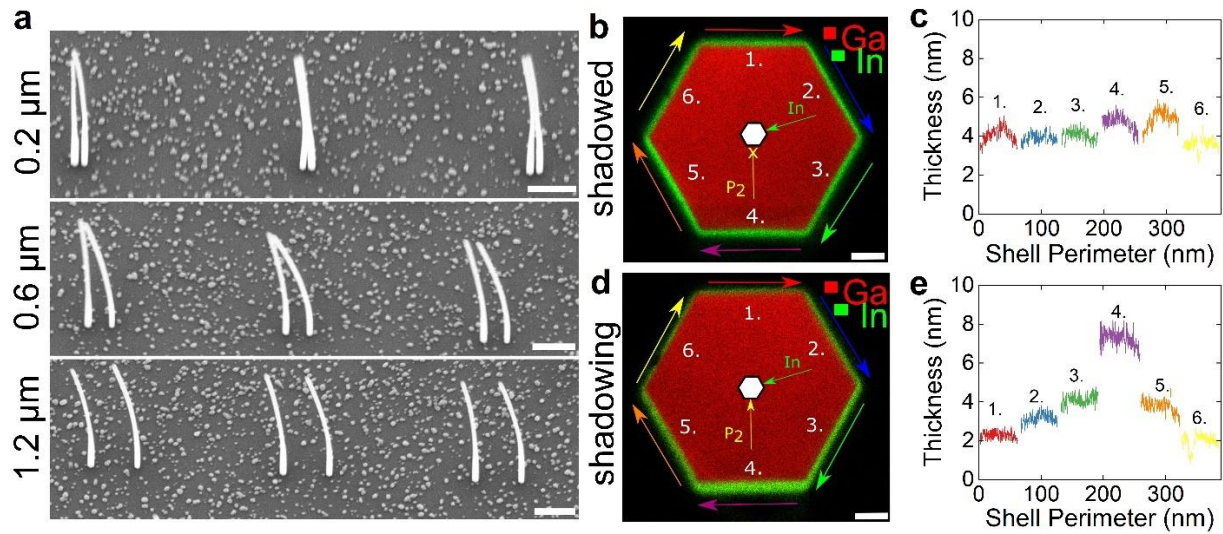

**Figure 3.** Nanowire pairs. (a) SEM images of nanowire pairs with shells grown at 440 °C imaged at a tilt of 30°. The pairs are aligned in the P$_2$ beam direction (incident from the right). Pairs are spaced by 0.2 µm, 0.6 µm and 1.2 µm. Scale bars for (a) are 1 µm. (b, d) EDS maps of a microtome cross-section of a shadowed nanowire (b) and shadowing nanowire (d) of a pair separated by 0.6 µm with shells grown at 380 °C. (c, e) Shell thickness plots corresponding to (b) and (d). The direction of the direct P$_2$ and In fluxes are indicated on the EDS maps. The yellow 'x' in (b) indicates that the direct P$_2$ flux is blocked from reaching the nanowire. The color and numbering of the EDS line scans corresponds to the labels on the EDS maps, indicating the path around the nanowire shell. Scale bars are 20 nm for the EDS maps.

The shell growth temperature was 440 °C for a planar deposition of 9 nm. For each pair, both nanowires are exposed to the same In beam incident at an azimuthal angle of 108° from the P$_2$ beam. For these growth conditions, we observe that pairs spaced by 0.2 and 0.6 µm can contact (Figure 3a). This is a consequence of the decrease in bending from the shadowed nanowire of the pairs, observed for these spacings. The shadowed nanowire experiences less asymmetric growth from the obstruction of the P$_2$ beam—in the ideal case of perfect shadowing, only receiving the uniform scattered P$_2$ flux on all sidewall facets. The efficacy of pair connection is strongly related to spacing. For pairs spaced by 0.2 µm, 86% of the observed pairs were found to be connected. The connection efficacy decreases to 24% for pairs spaced by 0.6 µm, as the nanowire alignment must be precise to result in connection for further spaced pairs. Pairs greater than 1.2 µm do not contact after bending or exhibit decreased bending due to shadowing (Figure 3a). As the pair spacing increases, less of the nanowire is shadowed (only the lower portion). For an incident P$_2$ inclination angle of $\theta$ (33.5° here), no part of the nanowire will be shadowed if the spacing is > $L \tan \theta$, where $L$ is the nanowire length.

Microtome nanowire cross-sections of nanowire pairs were characterized by TEM and EDS. Figure 3b–c shows EDS maps of a nanowire pair with a shell growth temperature of 380 °C and pair separation of 0.6 µm. The actual separation distance observed in TEM was 0.35 µm—a consequence of the nanowires being bent toward one another and suggesting the microtome slice was taken from near the mid-section along the nanowire length. The unshadowed nanowire (Figure 3c) exhibits a shell thickness distribution that is nearly identical to the isolated nanowire in Figure 2e–f, with the shell growth occurring mostly on the P$_2$-facing facets in the manner discussed earlier. This is expected as the unshadowed nanowires experience the same incident In and P$_2$ fluxes. In contrast, the shadowed nanowire of the pair in Figure 3b exhibits a highly symmetric shell of approximately 5 nm thickness. This is entirely consistent with the six sidewalls receiving only the symmetric scattered P$_2$ flux—the direct beam being shadowed. These results are fully consistent with the above result that the relative growth rate is proportional to the total incident P$_2$ flux on each facet. Furthermore, the total shell area measured from the cross-sections is nearly equal for both nanowires, 1420 ±190 nm² for the unshadowed nanowire and 1490±200 nm² for the shadowed nanowire, which is expected as the total shell growth is limited by the In flux for these group-V-rich growth conditions, which is the same on both nanowires. Thus, the assumption of a symmetric phosphorus scattering flux is experimentally confirmed by the shadowing of the P$_2$ beam and resulting symmetric shell.

Lastly, we explore the connection between nanowire pairs. Figure 4 displays TEM micrographs and corresponding diffraction patterns of two pairs initially separated by 0.2 µm connected during shell growth at 380 °C.

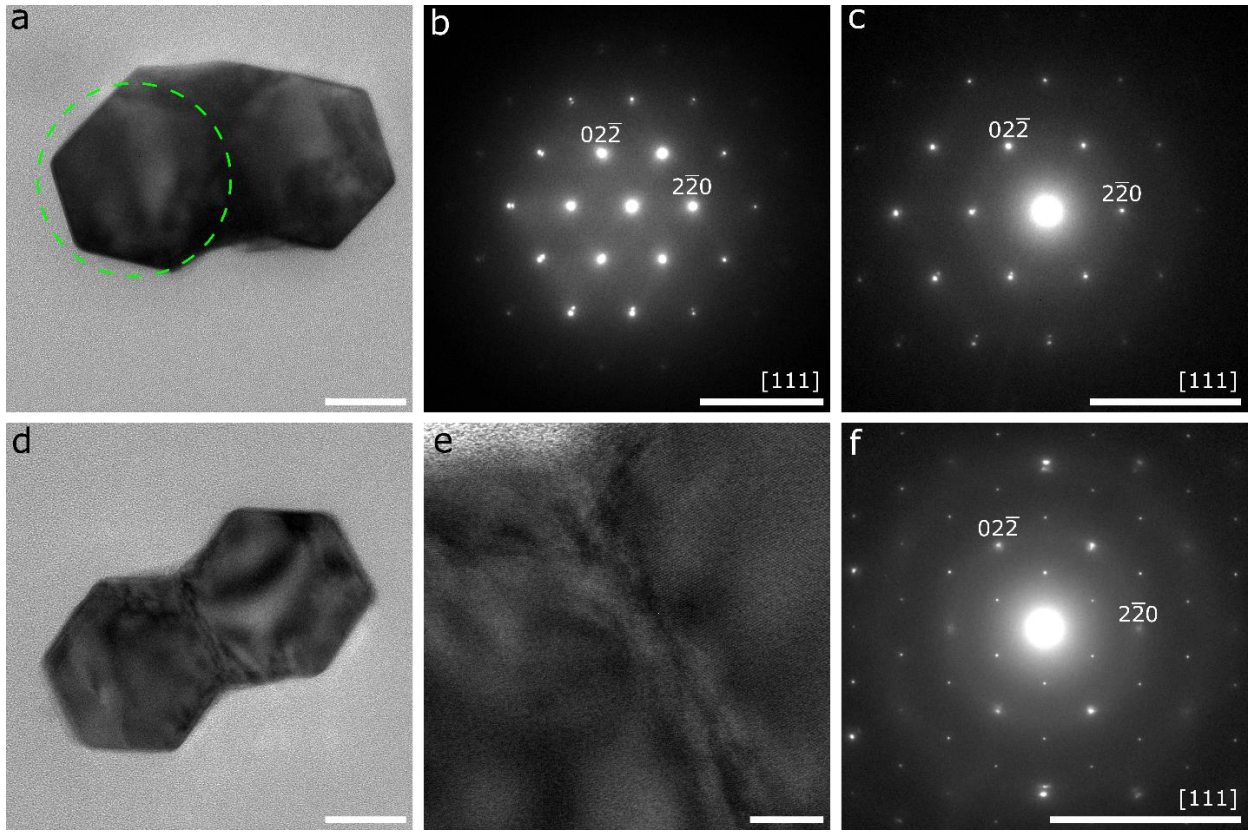

**Figure 4.** TEM characterization of fused nanowire pairs. (a) Bright-field micrograph of a connected pair. The corresponding diffraction pattern is shown in (b) and a selective-area diffraction pattern of the area indicated in (a) by the green dashed circle is displayed in (c). (d–f)

A second fused nanowire pair, where (e) shows a high-magnification image of the interface between the pairs in (d). (f) Diffraction pattern corresponding to (d). Scale bars are 50 nm in (a, d), 10 nm$^{-1}$ in (b, c, f) and 10 nm in (e).

For both nanowire pairs, the nanowires appear to be intimately fused together. The selective-area diffraction pattern of the left member in figure 4b shows a zinc-blend structure, and Figure 4c shows the complete pair pattern and contains faint rings indicative of amorphous structure. The amorphous growth is believed to occur at the fused interface of the two wires. Figure 4d shows a high-resolution TEM (HRTEM) of a second fused pair, with a higher magnification of the connection region shown in Figure 4e. The boundary between the fused pair exhibits as a dark contrast. Across the boundary, the crystal structure is misaligned as seen by the HRTEM (Figure 4e), and by the presence of additional spots in the diffraction pattern in Figure 4f. The diffraction pattern is aligned in the <111> direction with the right nanowire of the pair. Amorphous rings are also observed.

In summary, the symmetry and thus bending of nanowires with asymmetric lattice-mismatched shells is critically dependent on the adatom diffusion on the nanowire sidewalls during shell formation. InP shell growth was studied under two regimes: a low temperature/In-diffusion regime, where growth favors facets receiving direct In impingement, and a high temperature/In-diffusion regime, where the growth on a facet is linearly proportional to the incident P$_2$ flux—comprising the directional source flux and a symmetric re-evaporation flux (approximately 50% of the direct source flux). These results are consistent with established planar GaAs growth kinetics and have important implications for nanowire shell growth and prospective nanowire devices. The group-V-controlled growth regime was employed to quantify nanowire pair shadowing and to bend nanowires together to form connected arches, structures which are of high interest for nanowire chemical sensors and interconnects. Connected nanowires were found to form an intimate contact, which is highly encouraging for electrical conductivity between these structures and related prospective devices. This detailed understanding of 3D nanoepitaxy elucidates important processes which can be employed in fabrication of novel 3D nanostructures of other materials and will help pave the way for bottom-up, scalable fabrication of nanowire sensors based on bent nanowires.

**METHODS**

Samples were grown by gas-source MBE on patterned $SiO_2$-covered Si(111) substrates (substrate preparation described in the Supporting Information). The Ga and In fluxes were provided by solid-source effusion cells, while the $P_2$ flux was provided via phosphine flow cracked at 1000 °C. The sources were incident on the substrate at an angle of 33.5° from the substrate normal. GaAs nanowire cores 4.6 μm long and diameter tapering from 180 nm to 100 nm base to tip were grown as described previously.[3] After core growth, the substrate rotation angle (and thus the angle of the incident fluxes with respect to the nanowire side facets) was set to a fixed position for InP shell deposition. InP shells were deposited under a V:III flux ratio (P:In) of 10:1 at an In flux corresponding to a planar InP growth rate of 0.25 μm/h. The In and $P_2$ fluxes were separated by an azimuthal angle of 108°. Nanowire pairs were aligned either in the direction of the In beam or the $P_2$ beam. InP shells were grown at various substrate temperatures: 210 °C, 295 °C, 380 °C, and 440 °C.

The nanowire morphology was examined by scanning electron microscopy (SEM) with a JEOL JSM-7000F and with TEM in a Talos F200X. Nanowire cross-sections were obtained from microtomy with a Leica UCT ultramicrotome. The microtome cuts were placed on TEM grids for imaging along the <111> nanowire axis with HAADF and HRTEM. EDS was performed in the TEM on the nanowire cross-sections. The shell thickness around the nanowire perimeter was deduced from the EDS maps (see Supporting Information).

ACKNOWLEDGMENTS

We thank Carmen Andrei for TEM support, Marcia Reid for microtome preparation, Chris Butcher for SEM support, Shahram Ghanad-Tavakoli for MBE support and Dr. Maureen J. Lagos for helpful TEM discussions. We are grateful to McMaster University's Centre for Emerging Device Technologies and University of Toronto's Toronto Nanofabrication Centre for providing facilities and nanowire fabrication technical support, and the Canadian Centre for Electron Microscopy. Our thanks to the Natural Sciences and Engineering Research Council of Canada for providing financial support under grant RGPIN-2020-05721.

# Phosphorus-Controlled Nanoepitaxy in the Asymmetric Growth of GaAs–InP Core–Shell Bent Nanowires

*Spencer McDermott, Trevor R. Smith, Ray R. LaPierre, and Ryan B. Lewis\**

*Department of Engineering Physics, McMaster University, L8S 4L7 Hamilton, Canada*

*\*Email: rlewis@mcmaster.ca*

This supporting information is added to give background to the measurement and analysis of the nanowire shell thickness distributions. The interpretation of the energy dispersive spectroscopy (EDS) is explained. Spectral information of a nanowire cross-section shows present elements in the microtome sample. The quantitative analysis of the InP shell EDS used to interpret shell thickness and distributions along nanowire side facets is explained. Lastly, the relationship between shell growth and $P_2$ flux is quantified.

**Cross-Section Spectrum from Energy Dispersive Spectroscopy**

The EDS spectrum was acquired from transmission electron microscopy (TEM) in the Talos F200X. The Talos 200X is outfitted with 4 in-column silicon drift detectors. The electron beam was accelerated at 200 kV. The scanning TEM (STEM) spatial resolution was 0.16 nm and EDS resolution was 1 nm. The spectrum acquired (Figure S1a) by EDS was fit in Thermo Fisher Scientific's Velox software (Figure s1b).

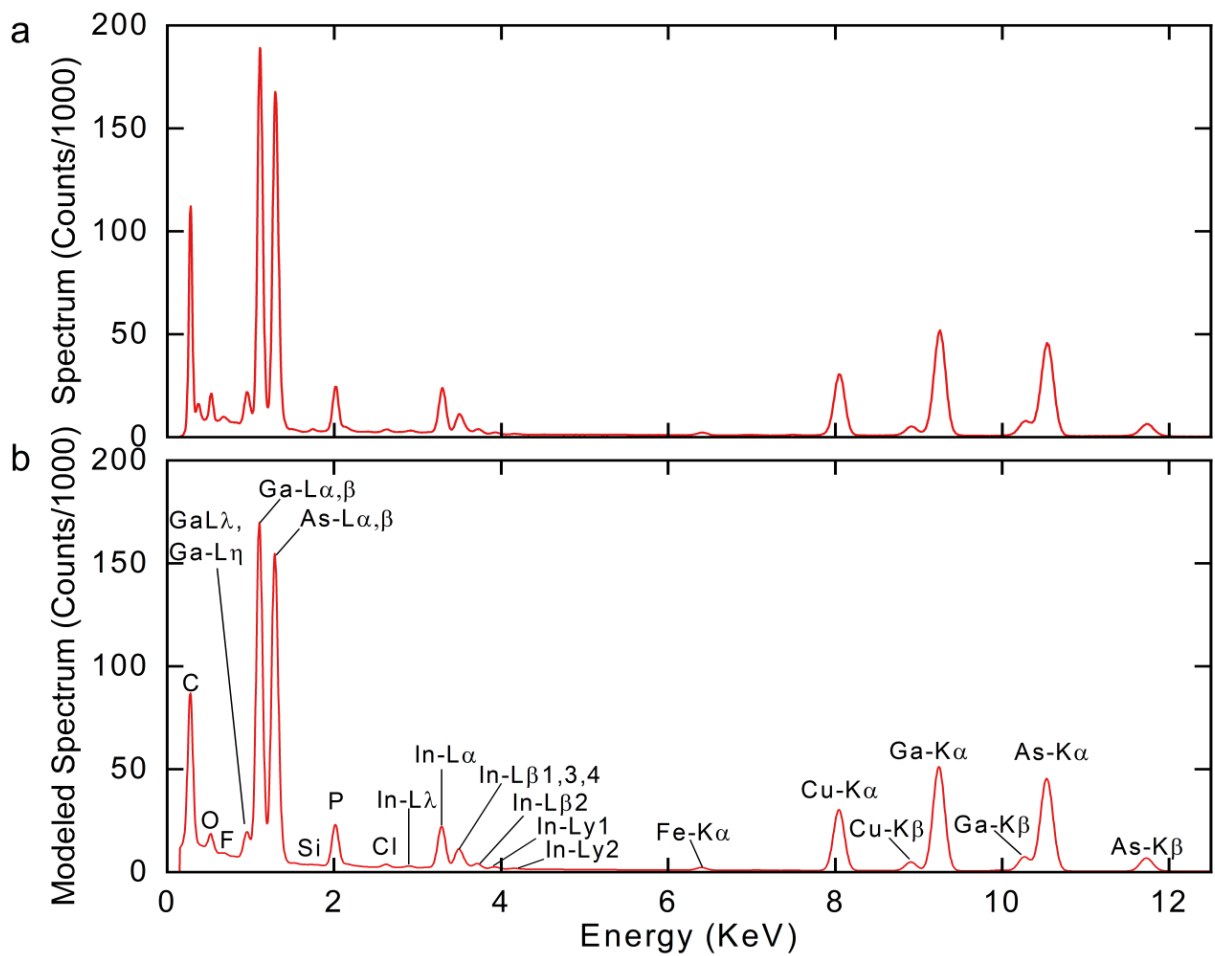

**Figure S1.** EDS spectrum of nanowire cross-section with shell grown a 210 °C. (a) The unmodified spectrum acquired in the Talos F200X. (b) A modeled spectrum with Velox to fit (a). Peaks are labeled with their corresponding emission.

The spectrum contains the elements of the III-V nanowire: In, Ga, P, and As peaks. There is a high number of counts of O and C from the Spurr's epoxy resin encasing the nanowire and from coating the grids in C to reduce charging during imaging. The other element with substantial counts is Cu from the Formvar-coated Cu grids used. The Si presence is suspected to be from contamination from the substrate during microtomy. The presents of F and Cl are expected to come from the Spurr's epoxy. Lastly, Fe is present as a background element in the microscope.

**EDS Shell Thickness Quantification**

For EDS mapping the characteristic emissions used are from the K-shell for Ga and the L-shell for In. The thickness of the shell can be measured directly from the EDS map and line scans along the side facets can reveal variation along the side facets. The intensity profile of the line scan is related to the thickness of the nanowire's microtome cross-section. This can be derived from the ζ-factor method for quantitative EDS:[1,2]

$$\zeta_m dI(x,y) = \rho_m T(x,y) C_m D_e dx dy \qquad \text{Eq. (1)}$$

$\zeta_m$ is a factor given to a material $m$ and TEM system, $I(x)$ is the intensity of the characteristic X-rays along the line scan for material $m$, $\rho_m$ is the density of the shell, $T(x,y)$ is the depth of material along the line scan, $C_m$ is the concentration of indium, and $D_e$ is the dose from the electron beam. This assumes that material composition is constant in the InP shell—as it is expected to be—and the microtome cut depth to be constant—as no wedging or other thickness variation of the nanowire core was observed. Thus, all variables remain constant over the EDS map where the shell is present, and the following relation holds true:

$$dI(x,y) = A(x,y) dx dy \qquad \text{Eq. (2)}$$

For In in the shell, $A(x,y)$ is nominally a 2D step function of the intensity of the L-shell characteristic emissions. The value is zero if the point $(x,y)$ is off the shell, or it is equal to the step maximum intensity ($A_{max}$) from Eq. (1) if the point is on the shell,

$$A(x,y) = \begin{cases} \frac{\rho_{In}T(x,y)C_{In}D_e}{\zeta_{In}} & (x,y) = on\ shell \\ 0 & (x,y) = off\ shell \end{cases} \quad \text{Eq. (3)}$$

Thus, for any line scan intensity along the nanowires side facet with scan width $W_{scan}$ along the scan length $l$:

$$I(l) = \int_{-W_{scan}/2}^{W_{scan}/2} A(w,l)dw \quad \text{Eq. (4)}$$

If the shell side facet is fully enclosed by the line scan width, the intensity is proportional to the shell thickness $T_{shell}(l)$ by the constant $A_{max}$:

$$I(l) = A_{max}T_{shell}(l) \quad \text{Eq. (5)}$$

This step function approximation neglects the convolution of the STEM probe as well as other uncertainties resulting from inelastic interactions that result in a gaussian profile. Thus, we fit the gaussian with a step function where the full width half maximum (FWHM) is taken to be the width of the step function and shell thickness ($T_{shell}$). In this way, the mean shell thickness ($\bar{T}_{shell}$) for an entire side facet is taken from a line scan shown in Figure S2a.

Supporting Information

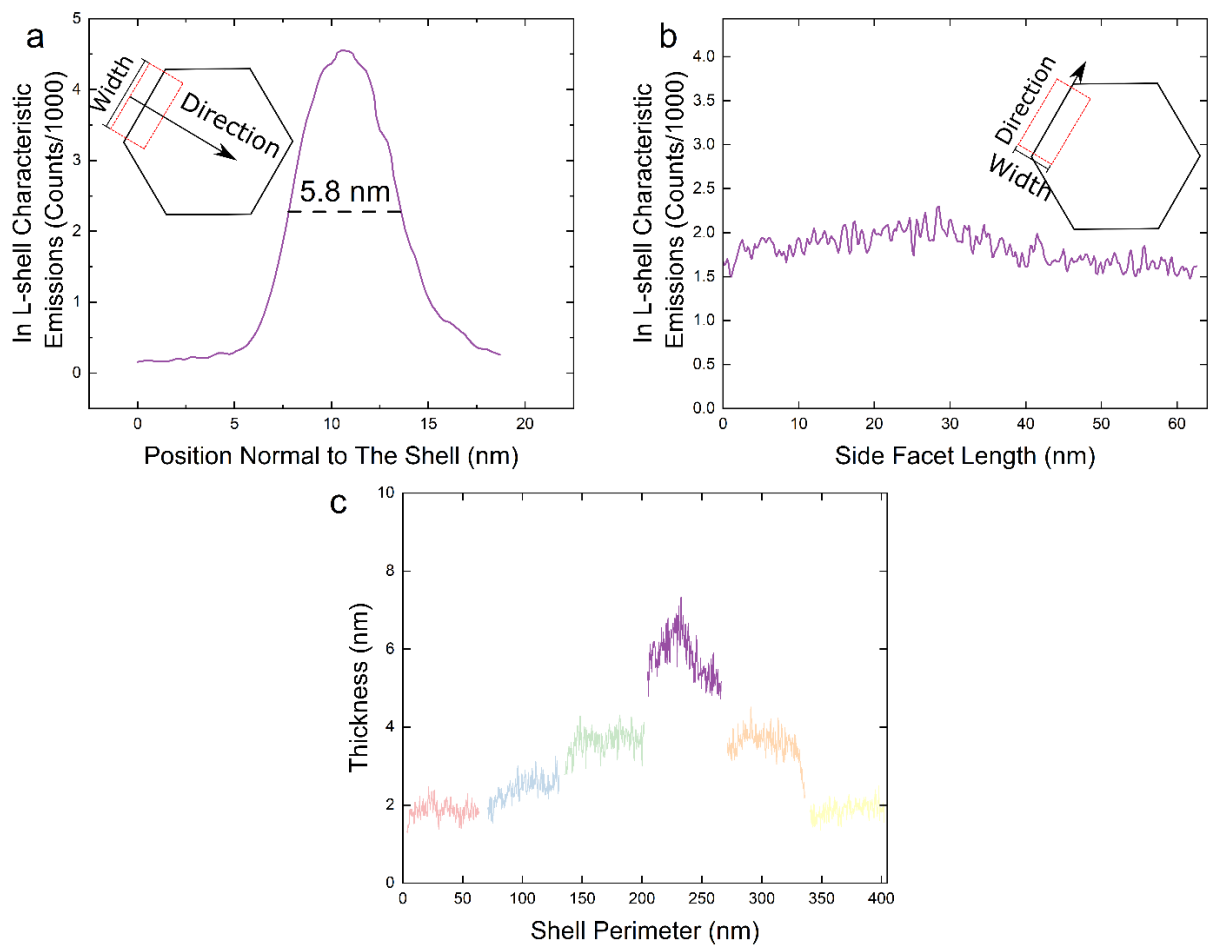

**Figure S2.** Shell thickness determination from EDS data. (a) A line scan normal to a side facet with a width that encompasses the side facet illustrated on the inset of the plot. The intensity profile is gaussian and the FWHM (5.8 nm) is taken to be the mean shell thickness $\bar{T}_{shell}$ for the side facet. (b) A line scan along the same side facet with width fully enclosing the shell as illustrated in the inset. The intensity profile along the side facet is plotted. (c) The shell thickness around the nanowire 6 side facets. The purple facet plotted is the side facet used in (a) and (b).

The mean facet thickness is compared to the mean intensity along the side facet (Figure S3b) to find the constant $A_{max}$

$$A_{max} = \frac{\bar{I}}{\bar{T}_{shell}} \qquad \text{Eq. (6)}$$

where $\bar{I}$ is the mean intensity along the side facet. The $A_{max}$ constant relates the intensity in Figure S2b of the side facet to the width along the side facet seen in Figure S2c as shown in Eq. (5). The same $A_{max}$ constant is used for all 6 side facets. For consistency and to minimize uncertainty, $A_{max}$ determined from the thickest shell facet is used for all scans. By using the same conversion factor for all intensity profiles of a single cross-section, it is insured that the intensity profiles are weighted relative to one another for comparison.

Relative weighting is effective for comparing side facets. However, it is noted that the nanowire cross-section grown at 210 °C shows significant bulging/mounding of the shell as seen in Figure 1a. Also, different intensities around the nanowire shell are observed in the EDS map, resulting in differences in intensity of the line scans and corresponding thicknesses that are not observed in the high-angle annular dark-field or EDS image. This contrast is expected to come from variations along the nanowire axis, resulting from the rough shell. For the cross-section with the shell grown at 210 °C, we use the weighting method described above for two reasons: first, the intensity does correspond to a shell material and second, to be consistent for comparison to the cross-section with shell growth at 380 °C. We stress that no variations in the core intensity are observed, only the shell grown at 210 °C.

## P₂ Impingement and Growth Rate

The growth rate for the side facets of the nanowires grown at 380 °C are taken from the mean value of the measured side facets:

$$\frac{\partial g_f}{\partial t} = \frac{\bar{T}}{\tau} \qquad \text{Eq. (7)}$$

$\frac{\partial g_f}{\partial t}$ is the growth rate, $\bar{T}$ is the measured thickness, and $\tau$ is the time of the shell deposition. For the six side facets from the nanowire in Figure 2f, we get the growth rates $0.16 \pm 0.03$ µm/h for facet 4, $0.10 \pm 0.02$ µm/h for facets 3 and 5, $0.05 \pm 0.01$ µm/h for facets 1 and 6, and $0.07 \pm 0.01$ µm/h for side facet 2.

The scattered flux is assumed to be equivalent on all the sidewall facets, which is confirmed by the symmetric shell for the shadowed nanowire in Figure 3c. We assume that the adatom incorporation diffusion length is linear[3–5], resulting in a linearly proportional growth rate from P₂ flux. Thus, the following equation results:

$$\frac{\partial g_f}{\partial t} = C\left[F_{P,direct}\langle \hat{b}_p \cdot \hat{n}\rangle + F_{P,scattered}\right] \qquad \text{Eq. (8)}$$

where $F_{P,direct}$ is the flux from the direct P₂ beam and $F_{P,scattered}$ is the P flux scattered from the substrate. $\langle \hat{b}_p \cdot \hat{n}\rangle$ is the geometric factor for the projected direct beam on a given facet. $\hat{b}_p$ is the normalized vector in the direct P₂ beam direction and $\hat{n}$ is the vector normal to the side facet. $C$ is the factor relating P₂ impingement to growth.

According to the facet numbering convention used in Figure 2e of the main text, facet 4 experiences the direct incident P₂ beam, with the normal of facets 3 and 5 having a projected angle of 60° from the direct P₂ flux. Therefore, the geometric factor $\langle \hat{b}_p \cdot \hat{n}\rangle$ (and thus the growth resulting from the direct P$_2$ flux) is expected to differ by a factor of $\cos(60°) = 0.5$ between facet 4 and facets 3 and 5. From the measured growth rates, the extracted growth resulting from direct P$_2$ impingement on facets 3 and 5 is found to be 45±21% and 43±20% of the direct growth on facet 4, respectively, in good agreement with the prediction based on the nanowire geometry.

**Substrate Preparation**

Samples were grown on Si(111) substrates covered by 30 nm of SiO$_2$ deposited by plasma-assisted chemical vapor deposition. The oxide layer was patterned by electron beam lithography (EBL) using a Raith EBPG 5000+ EBL system, followed by reactive ion etching. The pattern consisted of arrays of either individual or pairs of holes, spaced in a close-packed pattern separated by 5 or 10 µm in 100 x 100 µm fields. Hole pairs were spaced by 0.2, 0.6, 1.2, and 1.8 µm. Substrates were dipped in a solution of Fujifilm Buffered Oxide Etchant (10:1 NH4F:HF with Fujifilm surfactant) diluted with 9 parts water for 28 seconds immediately before being loaded into the MBE growth system.

Supporting Information